\documentclass[aps,prb,twocolumn,showpacs,tightenlines]{revtex4}
\usepackage{epsfig}
\usepackage{placeins}
\usepackage{psfrag}
\begin{document}
\def\Tr{\hbox{Tr}}
\def\ket#1{|#1\rangle}
\def\bra#1{\langle#1|}
\def\br#1{\langle#1}
\def\ketu#1#2{|#1\rangle^{#2}}
\def\brau#1#2{{}^{#2}\langle#1|}
\def\bru#1#2{{}^{#2}\langle#1}
\def\keti#1{\ketu{#1}{in}}
\def\brai#1{\brau{#1}{in}}
\def\bri#1{\bru{#1}{in}}
\def\keto#1{\ketu{#1}{out}}
\def\brao#1{\brau{#1}{out}}
\def\bro#1{\bru{#1}{out}}
\def\BCS{\text{BCS}}
\def\ketB{\ket{\BCS}}
\def\braB{\bra{\BCS}}
\def\expec#1{\langle\ #1\ \rangle}
\def\norm#1{\br{#1}\ket{#1}}
\def\proj#1{\ket{#1}\bra{#1}}
\def \ad {a^\dagger}
\def\kv{{\vec k}}
\def\lv{{\vec l}}
\def\pv{{\vec p}}
\def\qv{{\vec q}}
\def\xv{{\vec x}}
\def\yv{{\vec y}}
\def\measf#1{{\frac{d^3 #1}{(2\pi)^{3/2}(2\omega_{#1})^{1/2}}}}
\def\meas#1{{\frac{d^3 #1}{(2\pi)^{3}(2\omega_{#1})}}}
\def\measq#1{{\frac{d^4 #1}{(2\pi)^{4}}}}
\def\d#1#2{{\frac{\partial #1}{\partial #2}}}
\def\D#1#2{{\frac{d #1}{d #2}}}
\def\funcd#1#2{{\frac{\delta #1}{\delta #2}}}
\def\dn#1#2#3{{\frac{\partial^#3 #1}{\partial #2^#3}}}
\def\dt#1#2{{\frac{\partial^2 #1}{\partial #2^2}}}
\def\dT#1#2#3{{\frac{\partial^2 #1}{\partial #2 \partial #3}}}
\def\dTh#1#2#3#4{{\frac{\partial^3 #1}{\partial #2 \partial #3 \partial #4}}}
\def\dalem{{{\boxed{{ }^{ }}}}}
\def\del{\vec\nabla}
\def\pd{\partial}
\def\ulv{{\underline{\lv}}}
\def\ukv{{\underline{\kv}}}
\def\sl#1{{#1}\!\!\!/}
\def\dag{\dagger}
\def\J{{\bf J}}
\def\f{{\bf \phi}}
\def\F{{\bf \bar \phi}}
\def\P{{\bf p}}
\def\Q{{\bf q}}
\def\xp{x^\prime}
\def\r{{\bf r}}
\def\bt{{\bf t}}
\def\btp{{\bf t}^\prime}
\def\db{{\bf d}}
\def\rp{{\bf r}^\prime}
\def\s{{\bf s}}
\def\sp{{\bf s}^\prime}
\def\balpha{{\bf t}}
\def\plmi{\begin{array}{c}
+\\
-\\
\end{array}
}
\def \I {{\bf I}}
\def \q {{\bf q}}
\def \k {{\bf k}}
\def \kp {{\bf k^\prime}}
\def \p  {{\bf p}}
\def \pp {{\bf p^\prime}}
\def \ppp {{\bf p^{\prime\prime}}}
\def \pppp {{\bf p^{\prime\prime\prime}}}
\def \vsp {\vspace{-.2 in}}
\def \vspt {\vspace{-.2 in}}

\def \HB {H_{\text{BCS}}}
\def \sym {{sym}}
\def \el {\nonumber \\ &&}
%
\preprint{Preprint Numbers: \parbox[t]{45mm}{nucl-th/0305367}}
\title{BCS AND ATTRACTIVE HUBBARD MODEL COMPARATIVE STUDY}

\author{Nathan Salwen, Steven A. Sheets and Stephen R. Cotanch}

\affiliation{Department of Physics,
North Carolina State University, Raleigh,  NC 27695-8202}

\date{\today}
\begin{abstract}
  We extend previous studies of the BCS canonical approach for the
  attractive Hubbard model.  A derivation of the BCS formulation is
  presented for both the Hubbard and a simpler reduced Hamiltonian.
  Using direct diagonalization, exact one and two dimensional
  solutions for both Hamiltonians are compared to BCS variational
  calculations.  Approximate and exact ground state energies and
  energy gaps are computed for different electron number systems as
  well as correlation observables not previously predicted.
  Reproducing published one dimensional findings, the BCS is an
  excellent approximation for the Hubbard ground state energy but not
  energy gap, a finding that remains true in two dimensions.
  Propagators and correlators are found more sensitive to
  wavefunctions and appreciable differences are computed with the
  Hubbard model exhibiting a weaker degree of superconductivity than
  the BCS.  However for the reduced Hamiltonian model the BCS is an
  excellent approximation for all observables in both one and two
  dimensions.
\end{abstract}
\pacs{PACS number(s): 74.20.Fg, 71.24.+q, 71.10Fd, 71.10Li}
%
%
\maketitle

\section{
Introduction}
%



\psfrag{rho}{$\rho$}
\psfrag{Ne}{$N_e$}
\psfrag{Delta}{$\Delta_{N_e}$}
\psfrag{Cs}{$C_s$}
\psfrag{Cpx}{$C_{p_x}$}
\psfrag{Cdx2y2}{$C_{d_{x^2-y^2}}$}

\vsp

Since the advent of the theory of superconductivity,  developed by Bardeen,
Cooper and Schrieffer
[BCS]~\cite{bcs} and their contemporaries, Bogoliubov~\cite{bog} and
Valatin \cite{val},
countless investigations with these formalisms have been performed
in a variety of fields.  This widespread application of the BCS
approach is due not
only to its
power and utility but also because superior approximate or exact
solutions are generally not available.  This is especially true in
nuclear~\cite{dmp} and hadronic
structure~\cite{flsc} where physical insight has been limited by
uncertainties due to
approximations. Hence, further assessing the validity of
the BCS approximation is an
important endeavor
with broad attending interest.  In particular, there have been several studies
\cite{mar,bd,tmar,fmh} of the BCS approximation within the Hubbard model for which exact
solutions, such as those
provided by the Bethe Ansatz~\cite{lw}, are known but only in one
dimension.  The purpose of
this paper is to extend these investigations in several new directions: 1)
assess the BCS approximation
for not only the Hubbard but also a reduced Hubbard-like Hamiltonian;
2) in addition to energies and energy gaps also examine propagators and
correlation
observables; 3) evaluate the BCS accuracy in two dimensions.

In discussing the Hubbard model
it is natural to restrict ourselves to a fixed filling because the
Hamiltonian conserves
electron number.  Therefore we utilize the canonical BCS formulation with a
fixed number of
electrons.  This is
superior to the grand canonical approach which only constrains the average
electron number and also produces a higher ground state energy. A previous
pioneering study~\cite{mar}
notably detailed the excellent ground state energy agreement between the
exact Hubbard model and the
approximate canonical BCS solution.  However that work also determined,
depending upon electron
density, that the Hubbard and BCS energy gaps do deviate,
again in one dimension.
Related, an odd filling BCS application~\cite{tmar} in one dimension   also
found somewhat less precise
results.  Because the BCS approximation is conjectured to improve in higher
dimension~\cite{mar}, it is important
to perform a numerical assessment in two dimensions.

When computing with the BCS ground state many of the terms in
the Hubbard Hamiltonian, $H$, become irrelevant since only components which are
either diagonal in momentum space or create and destroy momentum modes
in pairs contribute.  Truncation to these remnant terms defines
the reduced Hamiltonian, $H_r$, which is specified in the next section where
the BCS Hubbard ground state energy, $E_{BCS}$, is found degenerate with
the BCS reduced
Hamiltonian value. Because $H$ contains further terms describing additional
dynamics we a
priori expect
$E_{BCS}$ to better approximate the  exact reduced Hamiltonian ground state
energy,
${E}_r$, than the exact Hubbard ground state energy, $E$.  Indeed, as
we detail below $E_{BCS}$ is an order of magnitude closer to  $E_r$
than  $E$ even though the variational treatment is very good for both.
Consequently a more complete BCS assessment, especially for systems modeled
by $H_r$,
should address other observables which, in contrast to energies, are more
sensitive to wavefunction details.  Accordingly, we also
calculate electron pair propagators, wavefunction inner products
and  $s$, $p$ and $d$-wave
correlators which are important for establishing superconductivity.

This paper is divided into four sections.  In section II the Hubbard and
reduced
Hamiltonians are introduced and a new formulation of the canonical BCS
method is
derived.
This section also  details formulas for energy gaps, densities,
propagators and $s$, $p$ and $d$-wave
correlators.  Section III  compares and discusses numerical results for a
variety of electron level
fillings and different number of lattice sites.  Finally, conclusions are
summarized in section IV.

\section{\label{sec:intro} THEORETICAL FORMULATION}

\subsection{Model Hamiltonians}

We follow the literature notation, especially Refs.~\onlinecite{mar}
and \onlinecite{tmar}, and
recall
the full Hubbard Hamiltonian in momentum space
\begin{eqnarray}
 H = \sum_{\p\sigma} \epsilon_\p a^\dag_{\p\sigma}a_{\p\sigma} +
  \frac{U}{V}\sum_{\p \pp \q} a_{\p\uparrow}^\dag
  a_{-\p+\q\downarrow}^\dag
  a_{-\pp+\q\downarrow}
  a_{\pp\uparrow} \, .
\label{h}
\end{eqnarray}
The reduced Hamiltonian retains the non-diagonal ($\p \ne \pp$) terms  with
$\q=0$ and
diagonal ($\p = \pp$) terms
\begin{eqnarray}
  H_r =
  \sum_{\p\sigma} \epsilon_\p a^\dag_{\p\sigma}a_{\p\sigma} +
  \frac{U}{V}\sum_{\p \pp, \p \ne \pp} a_{\p\uparrow}^\dag
  a_{-\p\downarrow}^\dag
  a_{-\pp\downarrow}
  a_{\pp\uparrow} \nonumber
   \\ +
  \frac{U}{V}\sum_{\p \q} a_{\p\uparrow}^\dag
  a_{-\p+\q\downarrow}^\dag
  a_{-\p+\q\downarrow}
  a_{\p\uparrow} \, .
\label{hr}
\end{eqnarray}
In the above equations $a^\dag_{\p\sigma}(a_{\p\sigma})$ creates
(annihilates) an electron with momentum $\p$ and spin $\sigma =$
$\uparrow$ or $\downarrow$.  In one dimension the electron kinetic energy is
$\epsilon_\p = -2t \cos(\p)$, while in two dimensions,
 $\epsilon_\p = -2t
(\cos(p_x) + \cos(p_y))$, with hopping rate
$t$ (set to $1$ in this work) between nearest neighbors. The momenta
are restricted to
multiples of $\frac{2\pi}{L}$, where $L$ is the linear extent of the
system. The number of sites $V$ is $L$ in one dimension or $L^2$ in
two dimensions.
Note that the last term in Eq. (\ref{hr}) can be expressed in terms of
a sum over the dummy index ${\bf l}= -\p +\q$ yielding $\frac{U}{V}N_{e\uparrow}
N_{e\downarrow}$, the product of spin up, $N_{e\uparrow}$, and down,
$N_{e\downarrow}$, electron number operators for a system having $V$ total
lattice sites.  Since we restrict this discussion to the attractive
Hubbard model the potential coupling parameter $U$ is negative.
The reduced Hamiltonian, $H_r$, has been chosen to only contain
elements of $H$ which have non-zero expectation values in the BCS state defined
in the next section.

\subsection{BCS Approximation}

We adopt the  formulation of Ref.~\onlinecite{mar} but present
an alternate BCS derivation.
For a given Hamiltonian, the BCS approximation entails using
the unnormalized variational wavefunction
\begin{equation}
  \ket{BCS} = \prod_\k (1 + g_\k a_{\k\uparrow}^\dag
  a_{-\k\downarrow}^\dag) \ket{0} \ ,
\label{bcs}
\end{equation}
to minimize the ground state energy.
Here $g_\k$ are the variational parameters and
$\ket{0}$ represents the vacuum state with no electrons.
 Note that the above Hamiltonians conserve particle number, but the state
$\ket{BCS}$ includes a mixture of fock components with different particle
number.
Because we are interested in states with fixed fillings we organize
these components by electron number $N_e = 2\nu$ with $\nu = 0, 1, 2, 3...$
\begin{eqnarray}
  \ket{BCS}= \sum_\nu \ket{\Psi_{2\nu}}
\end{eqnarray}
\begin{eqnarray}
  \ket{\Psi_{2\nu}} =
  \sum_{
    s(\nu:)
  }
  \prod_{i = 1}^\nu
  g_{\k_i} a_{\k_i\uparrow}^\dag a_{-\k_i\downarrow}^\dag
  \ket{0} \ .
\label{bcsn}
\end{eqnarray}
Here and below the cryptic summation notation
$s(\nu:)$ is used to represent the sum over all unordered sets of
momentum coordinates, $\k_1,\k_2,\ldots \k_\nu$, with the restriction that all
of the
momenta are different.

For a physical system with $N_e$ electrons the
BCS ground state energy, $E_{BCS}$, is calculated by minimizing the
expectation value $E_{2\nu}$
\begin{equation}
 E_{2\nu} \equiv
\frac{\bra{\Psi_{2\nu}}H\ket{\Psi_{2\nu}}} {\norm{\Psi_{2\nu}}} \ ,
 \end{equation}
i.e. $E_{BCS} = \min \, {E_{2\nu}}$ where the minimization is achieved by
varying the $g_\k$
as detailed
below.

Before proceeding several points are in order.  First, in the grand canonical
BCS formulation the
ground state energy is obtained by minimizing $\frac{\bra{BCS}H\ket{BCS}}
{\norm{BCS}} $.  Using
Eq. (\ref{bcs}) one can show that the grand canonical is greater than
the
minimum
canonical energy.
Second, combining Eqs. (\ref{h},\ref{hr},\ref{bcsn})  directly yields
the degenerate result
\begin{equation}
\frac{\bra{\Psi_{2\nu}}H\ket{\Psi_{2\nu}}} {\norm{\Psi_{2\nu}}}
 = \frac{\bra{\Psi_{2\nu}}H_r\ket{\Psi_{2\nu}}} {\norm{\Psi_{2\nu}}} \ ,
\end{equation}
verifying, as stated above, that the BCS ground state energy is the same
for both
Hamiltonians. This is also true for the grand canonical BCS energy.  Third, the
successive operation of $H_r$ on $\ket{0}$ only generates states
having electrons paired just like the state $\ketB$  with opposite momentum
and spin,
called Cooper pairs.  In general, the exact Hubbard ground state will
contain additional electron correlations between such pairs and the BCS
approach assumes these correlations can
be ignored.

Now, we compute the norm of the fock state
\begin{eqnarray}
  \norm{\Psi_{2\nu}} =
  \sum_{ s(\nu:)}
   \prod_{i = 1}^\nu
   g_{\k_i}^2 \ ,
\end{eqnarray}
and the unnormalized kinetic energy
\begin{eqnarray}
  \bra{\Psi_{2\nu}} \sum_{p\sigma} \epsilon_\p a^\dag_{\p\sigma}a_{\p\sigma}
  \ket{\Psi_{2\nu}}
  =
  2\sum_\p \epsilon_\p   g_\p^2
  \sum_{s(\nu-1:\p)}
  \prod_{i}^{\nu - 1} g_{\k_i}^2 \ . \nonumber
\end{eqnarray}
The modified summation index $s(n:\p_1,\p_2\,\ldots)$ now
denotes the sum over all unordered sets of
momentum coordinates, $\k_1,\k_2,\ldots \k_n$, with the stronger 
restriction that all
momenta in the extended set, $\{\k_1,\ldots \k_n,\p_1,\p_2,\ldots\}$, are
different.
For the potential energy we first calculate
\begin{eqnarray}
\lefteqn{
  \sum_\p a_{-\p\downarrow}a_{\p\uparrow}\ket{\Psi_{2\nu}} } \nonumber\\
  &=&
  \sum_\p a_{-\p\downarrow}a_{\p\uparrow}
  \sum_{s(\nu:)}
  \prod_{i = 1}^\nu
  g_{\k_i} a_{\k_i\uparrow}^\dag a_{-\k_i\downarrow}^\dag
  \ket{0} \nonumber \\
  &=&  \sum_\p g_\p
  \sum_{s(\nu-1:\p)}
  \prod_{i = 1}^{\nu-1}
  g_{\k_i} a_{\k_i\uparrow}^\dag a_{-\k_i\downarrow}^\dag
  \ket{0} \ .
\end{eqnarray}
Then the unnormalized potential energy matrix element is
\begin{eqnarray}
\lefteqn{
  \bra{\Psi_{2\nu}}
  \frac{U}{V}\bigg(\sum_{\p} a_{\p\uparrow}^\dag
  a_{-\p\downarrow}^\dag
  \sum_{\pp \ne \p}
  a_{-\pp\downarrow}
  a_{\pp\uparrow} + N\uparrow N\downarrow \bigg)
  \ket{\Psi_{2\nu}}} \nonumber \\
  &=&
  \frac{U}{V}
  \bigg(\sum_{\p,\pp}
  g_\p g_{\pp}
  \sum_{
    s(\nu-1:\p,\pp)
  }
  \prod_{i = 1}^{\nu-1}
  g_{\k_i}^2
  + \nu^2 \norm{\Psi_{2\nu}}
  \bigg) \ . \nonumber
\end{eqnarray}
Note well that the sum over the set $s(\nu-1:\p,\pp)$ ensures $\p
\ne \pp$.
Now define
\begin{eqnarray}
  A_n^m(\p_1,\ldots \p_m) \equiv
  \sum_{
    s(n:\p_1,\p_2,\ldots \p_m)
  }
  \prod_{i = 1}^{n}
  g_{\k_i}^2 \ ,
\end{eqnarray}
and
\begin{eqnarray}
  B_n^m(\p_1,\ldots \p_m) \equiv   \frac{A_n^m(\p_1,\ldots
\p_m)}{A_{\nu}^0} \ ,
\end{eqnarray}
which both vanish if any two momenta are equal, to express $E_{2\nu{}}$ as
\begin{eqnarray}
  \lefteqn{E_{2\nu} - \nu^2\frac{U}{V}}   \\
  &=&
  \frac{
   2 \sum_\p \epsilon_\p g_\p^2 A_{\nu-1}^1(\p)
    + \frac{U}{V}\sum_{\p,\pp } g_\p g_{\pp}
    A_{\nu-1}^2(\p,\pp)
    }{A_{\nu}^0} \nonumber \\
   &=& 2\sum_\p \epsilon_\p g_\p^2 B_{\nu-1}^1(\p)
    + \frac{U}{V}\sum_{\p,\pp} g_\p g_{\pp}
    B_{\nu-1}^2(\p,\pp)  . \nonumber
\end{eqnarray}

Next, we evaluate the BCS variational equation for a state with
$N_e = 2\nu$ electrons
\begin{eqnarray}
  \d{E_{2\nu}}{g_\p} = 0 \ ,
\end{eqnarray}
by  first calculating
\begin{eqnarray}
  \d{A_n^m(\p_1,\ldots \p_m)}{g_\p} &=&
  2g_\p \sum_{
    s(n-1:\p_1,\p_2,\ldots \p_m,\p)
  }
  \prod_{i = 1}^{n-1}
  g_{\k_i}^2 \nonumber \\
  &=& 2g_\p A_{n-1}^{m+1}(\p_1,\ldots,\p_{m},\p) \ .
\end{eqnarray}
Then we use the product rule for differentiating,
\begin{eqnarray}
  \lefteqn{
    A_{\nu}^0\ \d{B_n^m(\p_1,\ldots \p_m)}{g_\p}
  } \\
  &=&
  \d{A_n^m(\p_1,\ldots \p_m)}{g_\p} -
  B_n^m(\p_1,\ldots \p_m) \d{A_{\nu}^0}{g_\p} \nonumber \\
 &=&
  2g_\p \big(A_{n-1}^{m+1}(\p_1,\ldots \p_m,\p) -
  B_n^m(\p_1,\ldots \p_m) A_{\nu-1}^1(\p)\big) \nonumber  ,
\end{eqnarray}
to obtain
\begin{eqnarray}
  \lefteqn{0 = A_\nu^0\ \d{E_{2\nu}}{g_\p} } \nonumber
  \\
  &=&
  4\epsilon_\p g_\p  A_{\nu-1}^1(\p)
  + 4\sum_{\pp} \epsilon_\pp g_\pp^2
  g_\p A_{\nu-2}^{2}(\pp,\p) \el
  + \frac{2U}{V}\sum_{\pp} g_\pp A_{\nu-1}^2(\pp,\p) \el
  + \frac{2U}{V}\sum_{\pp, \ppp} g_\pp g_\ppp
  g_\p A_{\nu-2}^3(\pp,\ppp,\p) \el
  - 2g_\p A_{\nu-1}^1(\p) (E_{2\nu} - \frac{\nu^2 U}{V}) \ .
\end{eqnarray}
Solving for $g_\p$ yields
\begin{eqnarray}
  g_\p = \frac{U}{V{{\cal{D}}_e}}
    \sum_{\pp} g_\pp A_{\nu-1}^2(\pp,\p) \ ,
\label{gp}
\end{eqnarray}
with the denominator,
\begin{eqnarray}
  \lefteqn{{\cal{D}}_e =     A_{\nu-1}^1(\p)
    (E_{2\nu} - \frac{\nu^2 U}{V} -2\epsilon_\p) }
  \el
  - 2\sum_{\pp} \epsilon_\pp g_\pp^2
  A_{\nu-2}^{2}(\pp,\p)
  \el
  - \frac{U}{V}\sum_{\pp, \ppp} g_\pp g_\ppp
  A_{\nu-2}^3(\pp,\ppp,\p) \ .
\end{eqnarray}
Equation (\ref{gp}) is equivalent to Eq. (34) of Ref. \onlinecite{tmar}  (note
their equation has a misprint
involving a missing $g_\p$).

In the case of odd filling, $N_e = 2\nu + 1$, the variational state has a
single specific unpaired,
say up,
electron with momentum $\q$, chosen to minimize
the energy:

\begin{eqnarray}
  \ket{\Psi_{2\nu+1}} =
  \sum_{
    s(\nu:\q)
  }
  \bigg(
  \prod_{i = 1}^\nu
  g_{\k_i} a_{\k_i\uparrow}^\dag a_{-\k_i\downarrow}^\dag
  \bigg)
  a_{\q\uparrow}^\dag   \ket{0} \ .
  \label{eq:BCSodd}
\end{eqnarray}
Operating with the Hamiltonian on this state
shifts the energy by $\epsilon_{\q\uparrow} + \frac{U}{V} N_\downarrow$
and repeating the above analysis yields
\begin{eqnarray}
  \lefteqn{E_{2\nu+1} -  \epsilon_\q - \nu(\nu+1)\frac{U}{V}}
  \nonumber \\
  &=&
  \frac{ 2\sum_\p \epsilon_\p g_\p^2 A_{\nu-1}^2(\p,\q)
    + \frac{U}{V}\sum_{\p, \pp} g_\p g_{\pp}
    A_{\nu-1}^3(\p,\pp,\q)
  }{A_{\nu}^1(\q)} \nonumber  \\
  &=&
  2\sum_\p \epsilon_\p g_\p^2 C_{\nu-1}^1(\p)
  + \frac{U}{V}\sum_{\p, \pp} g_\p g_{\pp}
  C_{\nu-1}^2(\p,\pp) \ ,
\end{eqnarray}
where
\begin{eqnarray}
  C_n^m(\p_1,\ldots \p_m) \equiv   \frac{A_n^{m+1}(\p_1,\ldots
\p_m,\q)}{A_{\nu}^1(\q)} \ ,
\end{eqnarray}
is equivalent to $B_n^m(\p_1,\ldots \p_m)$ defined on a space with the
$\q$ mode removed.  Finally, the odd electron system variational equation
reduces to
\begin{eqnarray}
  g_\p
  =
  \frac{U}{V{{\cal{D}}_o}}
    \sum_{\pp} g_\pp A_{\nu-1}^3(\pp,\p,\q) \  ,
\end{eqnarray}
where

\begin{eqnarray}
  \lefteqn{{\cal{D}}_o =     A_{\nu-1}^2(\p,\q)
    (E_{2\nu+1} - \frac{\nu(\nu+1) U}{V} -2\epsilon_\p - \epsilon_\q)}
  \el
  - 2\sum_{\pp} \epsilon_\pp g_\pp^2
  A_{\nu-2}^{3}(\pp,\p,\q)
  \el
  - \frac{U}{V}\sum_{\pp, \ppp} g_\pp g_\ppp
  A_{\nu-2}^4(\pp,\ppp,\p,\q) \ . \qquad\qquad
\end{eqnarray}

\subsection{Exact diagonalization}

Using PARPACK, a parallel implementation of the Lanczos algorithm, we
exactly calculated the Hubbard and reduced Hamiltonian ground states
as a function of the electron density.  The dimension of the Hilbert
space for a 16 site Hubbard model is $4^{16} \approx 4\times 10^9$.
Since the Hamiltonians conserve momentum and particle number we may
restrict to subspaces with fixed momentum and filling.  However, the
dimension for half-filling at $0$ momentum is still greater than
$10^7$.  There are about 2000 non-zero terms for each state, so the
Hubbard Hamiltonian matrix cannot be stored in memory even in a sparse
format.  Instead, matrix elements were generated at each step in the
Lanczos algorithm.
The
diagonalization took about 3 hours using 30\, 1.2 Ghz  Athlon
processors in parallel.

For  even $N_e = 2\nu$, we restrict to the zero momentum subspace which
yields the minimum energy eigenvector.  For odd $N_e$,
the momentum with lowest ground state energy is dependent on the
coupling $U$.  For $U=0$ it is obvious
that the lowest energy will have the momentum of the $\nu + 1$th
lowest energy mode.  For high $U$, with less than half filling,
the lowest
energy will have $0$ momentum, but for the value $U=-10$ used
throughout this paper, the momentum is the same as for $U=0$.

The total momentum of the ground state for odd $N_e$ corresponds to
occupation of a specific unpaired momentum mode.  For the Hubbard
model, occupation of this mode is not fixed by the action of the
Hamiltonian but, for the reduced model, all relevant basis states have
the same unpaired mode.  In this regard, the reduced model ground
state is the same as the BCS state which has the single unpaired mode
$\q$ (Eq. \ref{eq:BCSodd}).

\subsection{Observables and theoretical constructs}

To properly assess any approximation scheme, as well as investigate the
robustness of a
particular dynamic model, it is important to calculate and compare a variety of
observables and theoretical constructs.

In addition to the ground state energy, $E$, and energy density, $E/V$,
another important observable is
the single particle energy gap,
$\Delta_{N_e}$,
\begin{eqnarray}
  \Delta_{N_e} &=& \frac{1} {2} [E_{N_e + 1} - E_{N_e} + E_{N_e - 1} -
E_{N_e}] \nonumber \\
        &=& \frac{1} {2}
[E_{N_e + 1} - 2E_{N_e} + E_{N_e - 1}] \ .
\end{eqnarray}
In the grand canonical formulation, a quasi-particle  excitation is
produced by the action of $a_{\p\uparrow}^\dag$, or equivalently,
$a_{-\p\downarrow}$ on the state $\ketB$ ~\cite{schrieffer} with excitation
energy
\begin{eqnarray}
  \lefteqn{E_\p \equiv \frac{\braB a_{\p\uparrow}  H a_{\p\uparrow}^\dag
\ketB}{\norm{BCS}}
  -
  \frac{\braB  H \ketB}{\norm{BCS}}} \nonumber  \\
 &=& \sqrt{\Delta_{\BCS}^2 + \epsilon_\p^2} \ . \qquad\qquad\qquad
\qquad\qquad\qquad
\end{eqnarray}
Here $\Delta_{\BCS}$ is  the BCS gap and $\epsilon_\p$ is
dependent on the momentum $\p$.  The minimum excitation is non-zero
since $E_\p
\ge \Delta_{\BCS}$, thus accounting for the name ``gap''.  In the
canonical BCS formulation $a_{\p\uparrow}^\dag \Psi_{2\nu}$ is no
longer proportional to $a_{\p\downarrow} \Psi_{2\nu}$.  Therefore,
$\Delta_{N_e}$ averages the effects of adding and subtracting one
electron.


Because wavefunctions are more sensitive than energies to approximation
schemes, we also
address other operator wavefunction overlaps such as
propagators and
correlators~\cite{fmh}. Before introducing theses constructs we first specify
the generalized one,
$\rho_{\sigma \tau}(r,s)$, and two,  $\rho_{\sigma \sigma^{\prime} \tau
\tau^{\prime}
}(\r,\rp,\s,\sp)$,
particle density matrices for any model state
$\ket{\Psi}$ with unit norm
\begin{eqnarray}
  \rho_{\sigma \tau}(\r,\s)&=& \bra{\Psi}c_{\r\sigma}^\dag c_{\s\tau}\ket{\Psi}
\equiv
\expec{ c_{\r\sigma}^\dag c_{\s\tau} }
\end{eqnarray}
\begin{eqnarray}
\rho_{\sigma \sigma^{\prime} \tau \tau^{\prime}  }(\r,\rp,\s,\sp) &=& \expec{
c_{\r\sigma}^\dag
c_{\rp\sigma^{\prime}}^\dag c_{\s\tau} c_{\sp\tau^{\prime}} } \ .
\end{eqnarray}
 Here the indices $\r, \s$ represent possible electron lattice sites.
The position operators, $c_{\r\sigma}$ and $c_{\r\sigma}^\dag$,  are related
to the momentum operators by
\begin{eqnarray}
  \label{eq:pos1}
  c_\r &= \frac{1}{\sqrt{V}}\sum_\p e^{-i \p\cdot \r} a_\p
\end{eqnarray}
\begin{eqnarray}
  \label{eq:pos2}
  c_\r^\dag &= \frac{1}{\sqrt{V}}\sum_\p e^{i \p\cdot \r} a_\p^\dag \ .
\end{eqnarray}
Between states with the same filling
$\rho_{\uparrow \downarrow}(\r,\s) = \rho_{\downarrow\uparrow }(\r,\s) =
0$, while $\rho_{\uparrow \uparrow}(\r,\s)$ and $\rho_{\downarrow
  \downarrow}(\r,\s)$
are the one electron propagators for spin up or down, respectively.

Next we construct a variety of correlators
as the difference between a specific two particle density and the product
of one particle
densities.    For example, the on site
$s$-wave correlator,
$C_{s}(\r,\s)$, is
\begin{eqnarray}
C_{s}(\r,\s) &=& \rho_{\uparrow
\downarrow \downarrow \uparrow}(\r,\r,\s,\s) - \rho_{\uparrow
\uparrow}(\r,\s)\rho_{\downarrow \downarrow}(\r,\s) \ ,
\end{eqnarray}
which is the correlated electron pair propagator minus the product of two
(independent)
one electron propagators.  The extended
$s$-wave correlator, $C_{es}(\r,\s)$, is
\begin{eqnarray}
C_{es}(\r,\s) &=&\sum_{\bt, \btp}w_{es}({\balpha})
w_{es}({\balpha^{\prime}})
\bigg(\rho_{\uparrow
\downarrow \downarrow \uparrow}(\r,\r+\balpha,\s,\s+\balpha^{\prime})
\el
\qquad -
\rho_{\uparrow
\uparrow}(\r,\s)\rho_{\downarrow
\downarrow}(\r+\balpha,\s+\balpha^{\prime})\bigg)
\ ,
\end{eqnarray}
 which involves  a sum, ${\balpha} = (t_x , t_y )$, $\balpha^{\prime} =
(t_x^{\prime} , t_y^{\prime}
)$, over the four planar sites adjacent to lattice site $\r$ ($\s$) with
weights $w_{es}({\balpha}), w_{es}({
\balpha^{\prime}})$ given by $w_{es}(\balpha) = 1 $
for $\balpha = (1, 0), (0,1), (-1,0)$ or $(0, -1)$ and
$w_{es}(\balpha) = 0 $ otherwise.

We also list in Table I several other correlators, $C_{\sym}$, with
different symmetry
weights $w_{\sym}(\balpha)$ which are
important for investigating superconductivity.
See Ref.~\onlinecite{fmh} for
illustrations and further discussion.

\begin{table}[h]
  \centering
  \caption{Correlators}
  \begin{tabular}{|c|c|c|} \hline
\nopagebreak

    Correlator & $C_{\sym}$ & non-zero $w_{\sym}({\bf t})$ terms \\
\hline\hline
    $s$-wave & $C_{s}$ &
    $w_{s}(0,0)  = 1$ \\
    \hline
    $
    \begin{array}{c}
      \text{extended} \\
      \text{$s$-wave}
    \end{array}
    $
    & $C_{es}$ &
    $
    \begin{array}{c}
      w_{es}(1,0) = w_{es}(-1,0) = 1 \\
      w_{es}(0,1) = w_{es}(0,-1) = 1
    \end{array}
    $ \\
    \hline
    $\begin{array}{c}
      x \  \text{direction} \\
      \text{$p$-wave}
    \end{array}
    $
     & $C_{p_x}$ &
     $
    \begin{array}{c}
      w_{p_x}(1,0) = 1 \\
      w_{p_x}(-1,0) = -1
    \end{array}
    $
    \\
    \hline
    $d$-wave & $C_{d_{x^2 - y^2}}$ &
    $
    \begin{array}{c}
      w_{d_{x^2 - y^2}}(1,0) = w_{d_{x^2 - y^2}}(-1,0) = 1 \\
      w_{d_{x^2 - y^2}}(0,1) = w_{d_{x^2 - y^2}}(0,-1) = -1
    \end{array}
    $
    \\
    \hline
    $
    \begin{array}{c}
    \text{diagonal} \\
    \text{$d$-wave }
    \end{array}
    $
    & $C_{d_{xy}}$ &
    $
    \begin{array}{c}
      w_{d_{xy}}(1,1) = w_{d_{xy}}(-1,-1) = 1 \\
      w_{d_{xy}}(1,-1) = w_{d_{xy}}(-1,1) = -1
    \end{array}
    $
\\
\hline
  \end{tabular}
\end{table}

Since our system has translational symmetry, the correlator,
$C_{\sym}(\r,\r + \db)$, and propagator,
$\rho_{\sigma\sigma}(\r, \r+\db)$, do not depend on $\r$.  Therefore we can
write
\begin{eqnarray}
C_{\sym}(\db) \equiv C_{\sym}(\r,\r + \db) =
\frac {1} {V} \sum_{\s} C_{\sym}(\s,\s + \db) \\
\rho_{\sigma\sigma}(\db) \equiv \rho_{\sigma\sigma}(\r,\r + \db) =
\frac {1} {V} \sum_{\s} \rho_{\sigma\sigma}(\s,\s + \db)\ .
\end{eqnarray}

As discussed in Ref.~\onlinecite{fmh}, an effective measure of superconductivity
is the
concept of
Off-Diagonal Long Range Order
(ODLRO)  which can be
quantitatively characterized by the correlator or ODLRO function,
$C_{\sym}(\db)$.
The criterion for demonstrating superconductivity is that
$C_{\sym}(\db)$ remains finite,
positive and does not decrease to zero at large $\r$.

Using Eqs. (\ref{eq:pos1}, \ref{eq:pos2}),
the propagator can be written in momentum space as
\begin{eqnarray}
  \rho_{\uparrow\uparrow}(\db) &=&
  \frac{1}{V}\sum_{\r}
  \expec{ c^\dag_{\r\uparrow}c_{\r+\db\uparrow} } \nonumber \\
  &=&
  \frac{1}{V^2}  \sum_{
    \p,\q
    }
  P(\p,\q,\db)
  \expec{ a_{\p\uparrow}^\dag
  {a_{\q\uparrow}} }  \ ,
\end{eqnarray}
where
\begin{eqnarray}
  P(\p,\q,\db) &=&
  \sum_{r}
    e^{i(\p\cdot \r - \q\cdot (\r+\db))} \nonumber
    \\
    &=& V \delta_{\p \q}e^{-i\q\cdot \db} \ .
\end{eqnarray}
Thus
\begin{eqnarray}
 \rho_{\uparrow\uparrow}(\db)
  =
  \frac{1}{V}  \sum_{
    \p
  }
  e^{-i\p\cdot \db}
  \expec{
  a_{\p\uparrow}^\dag
  {a_{\p\uparrow}}} \ ,
\end{eqnarray}
and evaluating
$C_{\sym}(\db)$ in momentum space  yields
\begin{eqnarray}
  \lefteqn{C_{\sym}(\db) =  \frac{1}{V}
  \sum_{\r,\balpha,\balpha^\prime}
  w_{\sym}(\balpha) w_{\sym}({\balpha^\prime}) } \\ && \qquad \qquad
  \rho_{\uparrow\downarrow\downarrow\uparrow}(\r,\r+\balpha, 
  \r+ \db+ \balpha^\prime, \r + \db) \el
  =
  \frac{1}{V^3}
  \sum_{\p,\pp,\ppp,\pppp}
  P(\p,\pp,\ppp,\pppp,\db)
  \expec{a_{\p\uparrow}^\dag
  a_{\pp\downarrow}^\dag
  a_{\ppp\downarrow}
  a_{\pppp\uparrow}}  \qquad
  \\
  \lefteqn{
  P(\p,\pp,\ppp,\pppp,\db) =
  V\delta_{\p, \ppp + \pppp - \pp}
  e ^{-i\db \cdot (\ppp + \pppp)}}
\\ && \qquad\qquad\qquad
  \sum_{\balpha}
  w_{\sym}(\balpha)
  e^{i\pp\cdot \balpha}
  \sum_{\balpha^\prime}
  w_{\sym}({\balpha^\prime})
  e^{-i \ppp\cdot \balpha^\prime} \ . \nonumber
\end{eqnarray}
In particular for d-wave symmetry
\begin{eqnarray}
  \lefteqn{
  P(\p,\pp,\ppp,\pppp,\db) =
  4V\delta_{\p, \ppp + \pppp - \pp}
  e ^{-i\db \cdot (\ppp + \pppp)} }
  \el
  (\cos (p^\prime_{x}) - \cos(p^\prime_{y}))\
  (\cos (p^{\prime\prime}_{x}) - \cos(p^{\prime\prime}_{y}))\ .
\end{eqnarray}
The other correlators were calculated similarly.

\section{Numerical Results and Discussion}

\subsection{One dimensional systems}

\begin{figure}[htb]
  \begin{center}
   \epsfxsize=23pc \epsfbox{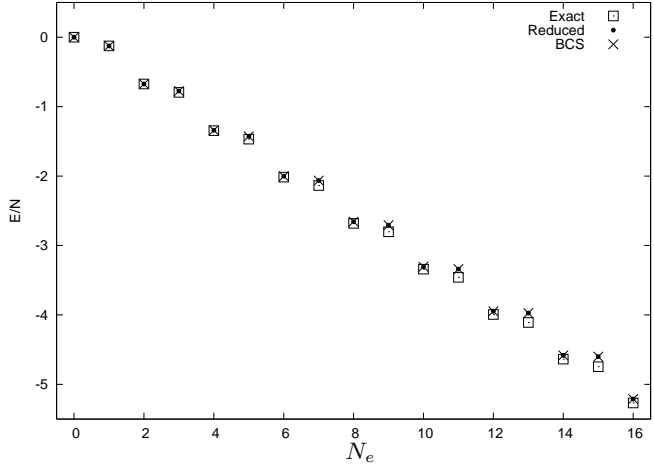}
   \caption{One dimensional ground state energy densities as a function of
     electron filling for the Hubbard (boxes) and reduced (circles)
     Hamiltonians.  Note the good agreement with the BCS results
     (crosses).}
  \label{fig:E1}
  \vsp
  \end{center}
\end{figure}

We first compute for a one dimensional lattice the approximate BCS and
exact Hubbard and reduced Hamiltonians ground state energies as a
function of level filling.  In Fig. \ref{fig:E1} the Hubbard (boxes),
$E/V$, reduced (circles), $E_r/V$, and BCS (crosses), $E_{BCS}/V$,
ground state energy densities are compared as a function of electron
filling, $N_e$, for $t = 1$, $V = 16$ sites and $U = - 10$.  The BCS and
exact Hubbard data reproduce the data in Refs.~\onlinecite{mar} and
\onlinecite{tmar}.  As they reported, the agreement is very good,
especially for the case of even filling.  The exact reduced data is
indistinguishable from the BCS data in the figure.  Over 99\% of the
difference between the BCS and the exact Hubbard is due to the
difference between $H_r$ and $H$ and only 1\% is due to quantum
correlations between different BCS pairs which are excluded by the
variational wavefunction.  Similar results have been obtained for
other potential strengths and different number of lattice sites (not
shown).

\begin{figure}[htb]
  \begin{center}
   \epsfxsize=23pc \epsfbox{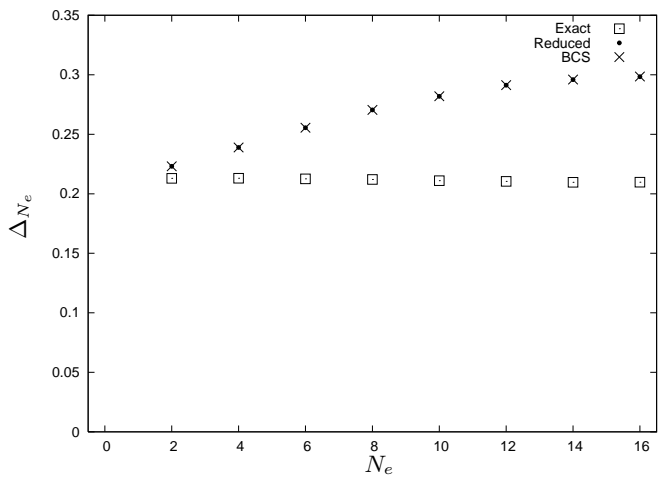}
   \caption{One dimensional energy gaps as a function of electron filling for the
     Hubbard (boxes) and reduced (circles) Hamiltonians.  The BCS
     (crosses) is least accurate near half-filling. }
   \label{fig:G1}
  \vsp
 \end{center}
\end{figure}

The Hubbard (boxes), reduced (circles) and BCS (crosses) energy gaps,
$\Delta_{N_e}$, are
compared in Fig. \ref{fig:G1}.  As with the ground state energy, the BCS gap is the
same for either the Hubbard or
reduced Hamiltonian.   In
contrast to energy density comparisons, the BCS-Hubbard difference is
more pronounced especially for half-filling, $N_e = V$.  The BCS is
still a good treatment for the reduced Hamiltonian.

\subsection{Two dimensional systems}

Having reproduced and extended one dimensional studies, we now address
two dimensional systems.  Figure 3 depicts the Hubbard, reduced and BCS
energy densities for
a two dimensional lattice having $V = 4 \, \text{x} \, 4 = 16$ sites.

\begin{figure}[htb]
  \begin{center}
   \epsfxsize=23pc \epsfbox{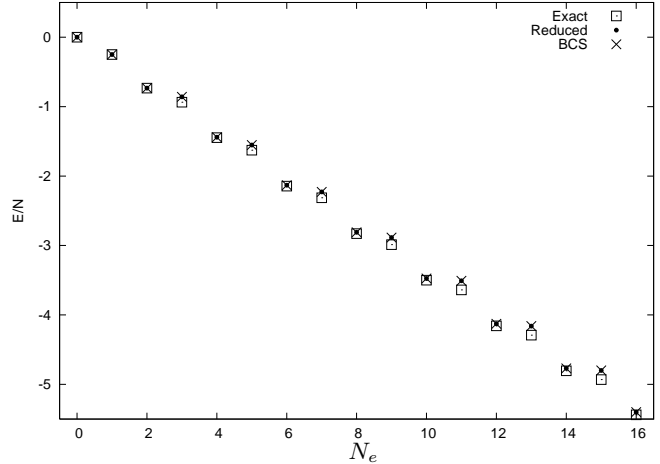}
   \caption{Two dimensional ground state energy densities as a function of
     electron filling for the Hubbard (boxes) and reduced (circles)
     Hamiltonians.  Note the good agreement with the BCS results
     (crosses).}
  \label{fig:E2}
  \vsp
  \end{center}
\end{figure}

\noindent
Again the BCS energies are close to the exact Hubbard eigenvalues and
indistinguishable from the reduced eigenvalues.
The corresponding
energy gap comparisons are displayed in Fig.  4. Contrary to
conjecture, the difference between the exact and BCS
energy gaps remains similar to the one dimensional case.

\begin{figure}[htb]
  \begin{center}
   \epsfxsize=23pc \epsfbox{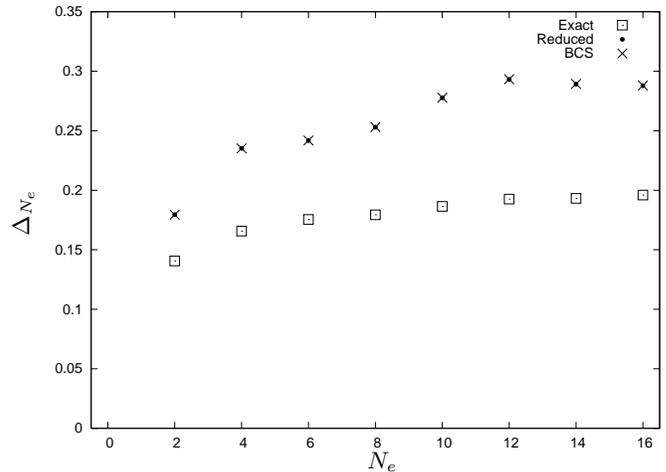}
   \caption{Two dimensional energy gaps as a function of electron filling for
     the exact Hubbard (boxes) and reduced (circles) Hamiltonians.
     Note the disagreement between the BCS (crosses) and the Hubbard energy
gaps, similar to the one dimensional result (compare to
     Fig. 2).}
 \label{fig:G2}
 \vsp
\end{center}
\end{figure}

We further note that the Hubbard and reduced energy eigenvalues are
generally in reasonably good agreement for all electron densities in
both one and two dimensions.  Accordingly, it would appear that the
reduced Hamiltonian incorporates the dominant physical dynamics.
This is fortuitous because $H_r$ can be solved more easily than $H$,
whether by the BCS approximation or by diagonalization, which in turn
facilitates investigating more complex systems in multi-dimensions.

\subsection{Wavefunction overlap}

\begin{figure}[htb]
  \centering
  \begin{center}
    \vspt
    \epsfxsize=20.5pc \epsfbox{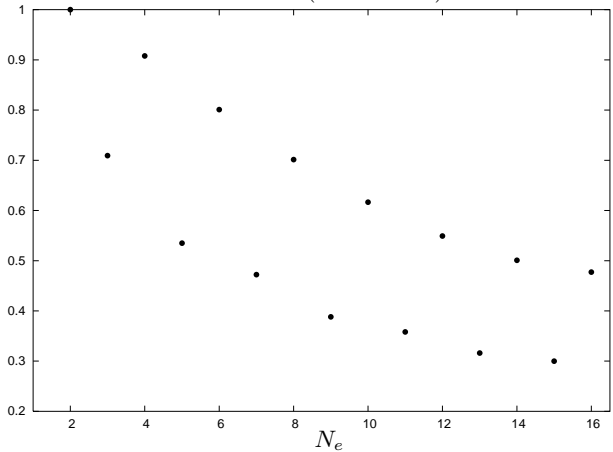}
    \caption{Inner product of the Hubbard and BCS vectors as a
      function of filling $N_e$ (one dimension).}
    \label{fig:ov1}
  \vsp
\end{center}
\end{figure}

\begin{figure}[htb]
  \centering
  \begin{center}
    \vspt
    \epsfxsize=20.5pc \epsfbox{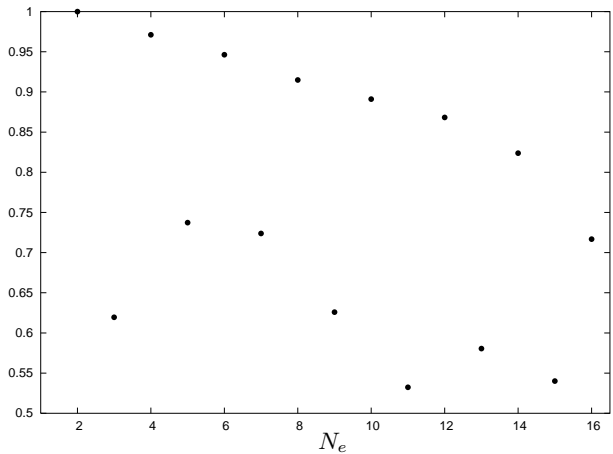}
    \caption{Inner product of the Hubbard and BCS vectors as a
      function of filling $N_e$ (two dimensions).}
    \label{fig:ov2}
  \vsp
\end{center}
\end{figure}

In general, a variational calculation  reproduces
the ground state energy much better than other observables.  It is therefore
of interest to examine
the overlap of the BCS and exact model wavefunctions.  Figure
\ref{fig:ov1} plots
the inner product of the exact, one dimensional Hubbard eigenvector with the
BCS state as a function of electron filling.  Notice the significant
deviation from
unity especially with increasing density.   As before, the BCS
and Hubbard results have the largest difference for systems near
half-filling and odd number of electrons.  In two dimensions the
overlap behavior is similar but now the values are
significantly higher (see Fig. \ref{fig:ov2}), consistent with the conjecture that
the BCS approximation improves with higher dimensionality.
The inner product of the reduced model eigenvector with the $\ket{BCS}$ is
$.999$ in
both 1 and 2 dimensions (not shown).

\subsection{Densities, propagators and correlators}

Returning to physical observables, this section compares
propagators and correlators for exact and approximate model wavefunctions.
Unless stated otherwise, all calculations were performed
for $U = - 10$, lattice sizes
$V = 16$ x $1$ and $V =4$  x $4$, and $N_e = 15$ and
$N_e = 16$.  Results for other fillings were also
calculated and found to be
qualitatively similar.

\begin{figure}[htb]
  \begin{center}
        \vspt
   \epsfxsize=23pc \epsfbox{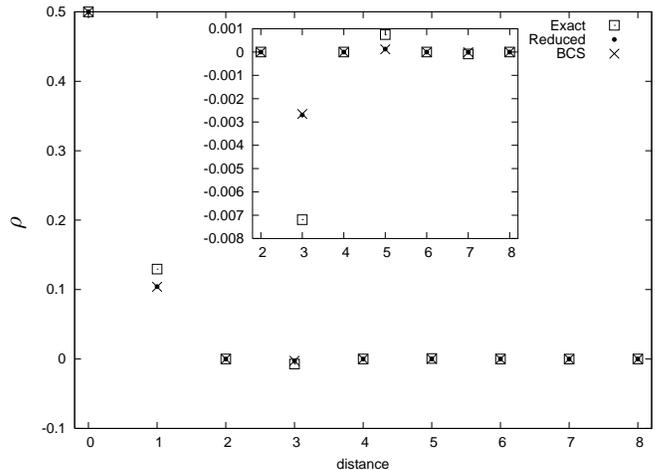}
   \caption{One electron propagators as a function of distance
(one dimension, $N_e = 16$).}
    \label{fig:p1_16}
  \vsp
\end{center}
\end{figure}

\begin{figure}[htb]
  \begin{center}
        \vspt
   \epsfxsize=23pc \epsfbox{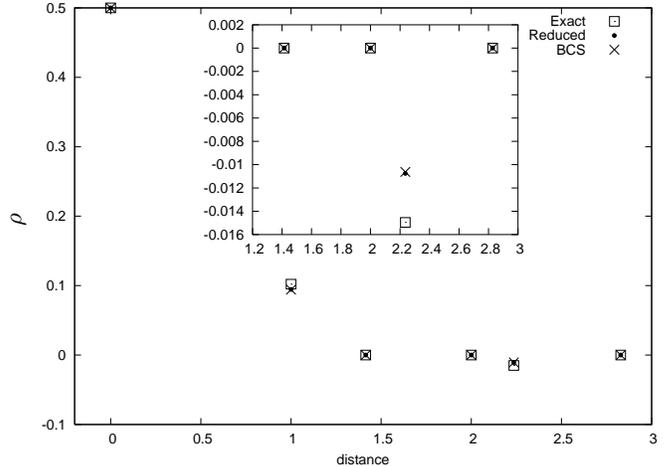}
\caption{One electron propagators as a function of distance for
(two dimensions, $N_e = 16$).}
    \label{fig:p2_16}
  \vsp
\end{center}
\end{figure}

Figures \ref{fig:p1_16} and \ref{fig:p2_16} compare the Hubbard,
reduced and BCS electron propagators, $\rho_{\uparrow\uparrow}(d)$, for
different lattice ranges $d$ and  $N_e = 16$ in one and
two dimensions, respectively.  The BCS and reduced propagators
are essentially identical and also similar to the Hubbard result.
At lower densities there are somewhat larger differences between the Hubbard
and BCS (or reduced) propagators.

\begin{figure}[htb]
  \begin{center}
        \vspt
    \epsfxsize=23pc \epsfbox{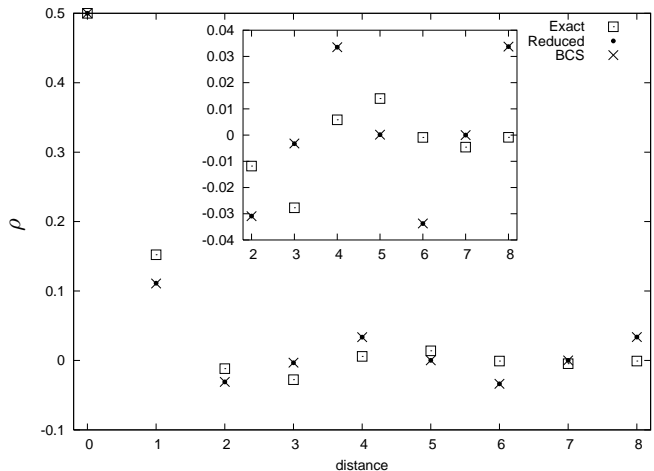}
\caption{One electron propagators as a function of distance
(one dimension, $N_e = 15$).}
    \label{fig:p1_15}
  \vsp
\end{center}
\end{figure}

\begin{figure}[htb]
  \begin{center}
        \vspt
   \epsfxsize=23pc \epsfbox{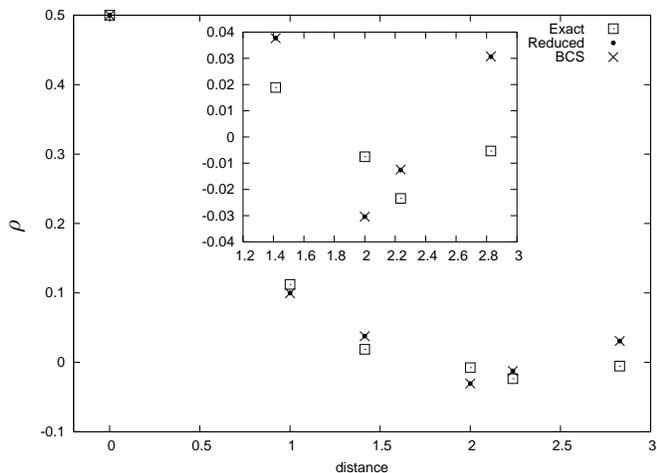}
   \caption{One electron propagators as a function of distance (two
     dimensions, $N_e = 15$).}
    \label{fig:p2_15}
  \vsp
\end{center}
\end{figure}

Figures \ref{fig:p1_15} and \ref{fig:p2_15}  compare the different model
single electron propagators again in one and
two dimensions, respectively, but now $N_e = 15$.  Notice that, just like
the model
energy comparisons, the Hubbard and reduced
propagators are more different for odd $N_e$, but again the BCS
accurately reproduces the reduced results.  Although lower $N_e$
Hubbard-reduced
comparisons yield even larger devations (not shown), overall  there
is no significant difference in agreement between the two propagators in
both one and
two dimensions.

\begin{figure}[htb]
  \begin{center}
        \vspt
   \epsfxsize=23pc \epsfbox{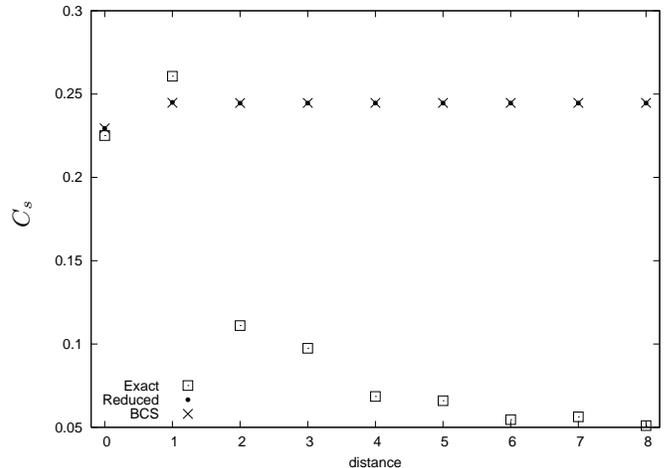}
\caption{The $s$-wave correlators as a function of distance (one
  dimension, $N_e=16$).}
    \label{fig:S1_16}
  \vsp
\end{center}
\end{figure}

\begin{figure}[htb]
  \begin{center}
        \vspt
   \epsfxsize=23pc \epsfbox{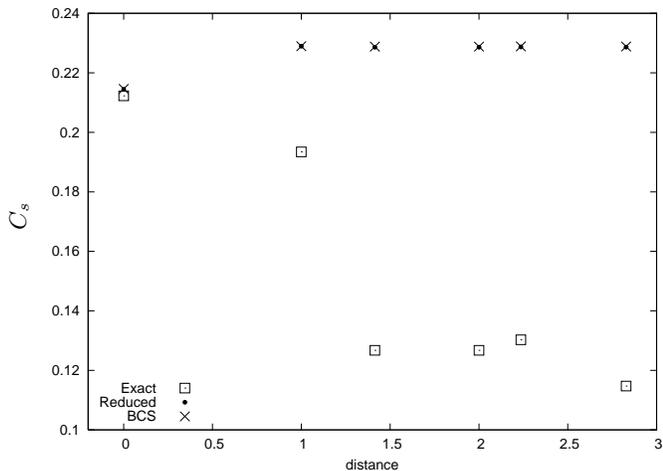}
\caption{The $s$-wave correlators as a function of distance (two
  dimensions, $N_e =
  16$)}
    \label{fig:S2_16}
  \vsp
\end{center}
\end{figure}

Figures \ref{fig:S1_16} and \ref{fig:S2_16} similarly compare the Hubbard,
reduced and BCS electron $s$-wave correlators, $C_s(d)$,
for $N_e =16$ in one and two
dimensions, respectively.  Now there are significant departures between the
Hubbard
and reduced results.  Notice that the reduced correlator has a larger 
magnitude
than the Hubbard value.  This is
consistent with the energy gap comparison, where the reduced gap was 
larger than
the Hubbard gap.  Also, but not shown, lower $N_e$
yields smaller Hubbard-reduced correlator differences.  
The BCS
again very accurately reproduces the reduced correlators.

The Hubbard and reduced $s$-wave correlators are closer in two
dimensions than in one dimension but the limited range of $d$ sampled
warrants caution in this conclusion.  By the Mermin-Wagner theorem
\cite{merminwagner}, the Hubbard model will not display long range
correlations at $T\ne 0$ in one or two dimensions.  However, according to the
Bethe ansatz\cite{solvable}, at least for large $|U|$, the ground
state energy of the one-dimensional system becomes part of a
continuous spectrum as the
system length increases.  The theorem does not apply at $T=0$ under
these circumstances and therefore long distance correlations are
possible.  The limited data shown in figures
\ref{fig:S1_16} and \ref{fig:S2_16} are inconclusive on the question of
long distance order.

Conversely, the BCS and reduced ODLRO functions are flat for increasing
$d$.  The Mermin-Wagner theorem does not apply to
the reduced Hamiltonian, even at $T\ne 0$, because it has long range
interactions.  For the ground state of the reduced Hamiltonian, the
$s$-wave correlator reduces to
\begin{eqnarray}
  \nonumber
  \frac{1}{V^2}
  \bigg(
  \sum_{\p , \pp}
  \expec{a_{\p\uparrow}^\dag
    a_{-\p\downarrow}^\dag
    a_{-\pp \downarrow}
    a_{\pp \uparrow}}
  \\
  +  \sum_{\p \ne \pp} 
  e^{-i \db \cdot (\p - \pp)}
  \expec{a_{\p\uparrow}^\dag
    a_{-\pp\downarrow}^\dag
    a_{-\pp \downarrow}
    a_{\p \uparrow}} \bigg).\qquad 
\end{eqnarray}
The second term decreases linearly with $V$ for large $\db$,
but the first term is independent of $\db$ and approaches a constant
as $V$ increases.  For $U,N_e,V \text{\ and\ }|d| \gg 1$, this
constant is $\frac{N_e(2V-N_e)}{4V^2}$.  A
precise limit can also be given for the BCS correlator at finite $V$.
Resummation of the correlator yields terms independent of $\db$,
\begin{eqnarray*}
  \sum_{\p \ne \pp} g_{\p} g_{\pp} B_{\nu-1}^2(\p,\pp)
  -
  \sum_{b=1}^\nu (-1)^b \sum_\p g_\p^{2b} B_{\nu-b}^1(\p),
\end{eqnarray*}
with additional terms decreasing rapidly to $0$ with increasing
$\db$.

\begin{figure}[htb]
  \begin{center}
        \vspt
   \epsfxsize=23pc \epsfbox{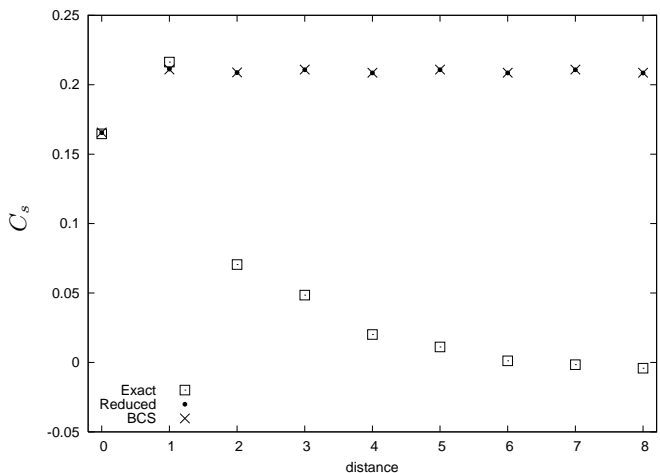}
\caption{The $s$-wave correlators as a function of distance (one dimension,
$N_e=15$).}
    \label{fig:S1_15}
  \vsp
\end{center}
\end{figure}

\begin{figure}[htb]
  \begin{center}
        \vspt
   \epsfxsize=23pc \epsfbox{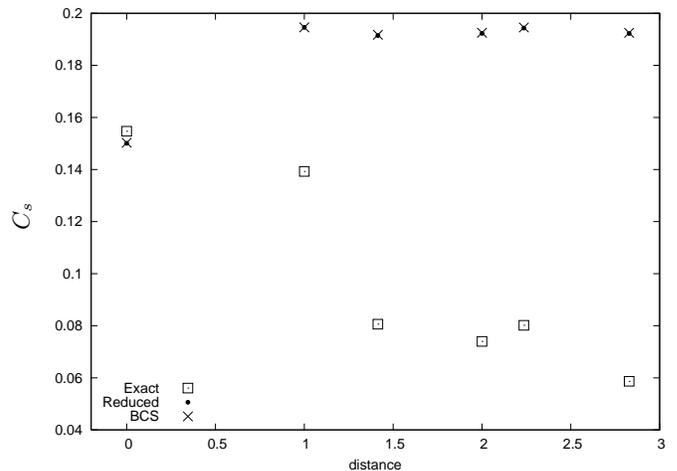}
   \caption{The $s$-wave correlators as a function of distance (two dimensions,
$N_e=15$).}
    \label{fig:S2_15}
  \vsp
\end{center}
\end{figure}

The Hubbard,
reduced and BCS electron $s$-wave correlators, $C_s(d)$, for $N_e =15$ are
compared
in Figs. \ref{fig:S1_15} and \ref{fig:S2_15} for one and two dimensions,
respectively.  The differences are more dramatic for odd fillings as
the Hubbard correlator in one dimension falls
rapidly to $0$.  The two dimensional Hubbard correlator also falls
rapidly within the limited range in $d$.  In contrast, the BCS and
reduced models, with a single specific unpaired electron mode,
still exhibit long range correlations.

\begin{figure}[htb]
  \begin{center}
        \vspt
   \epsfxsize=23pc \epsfbox{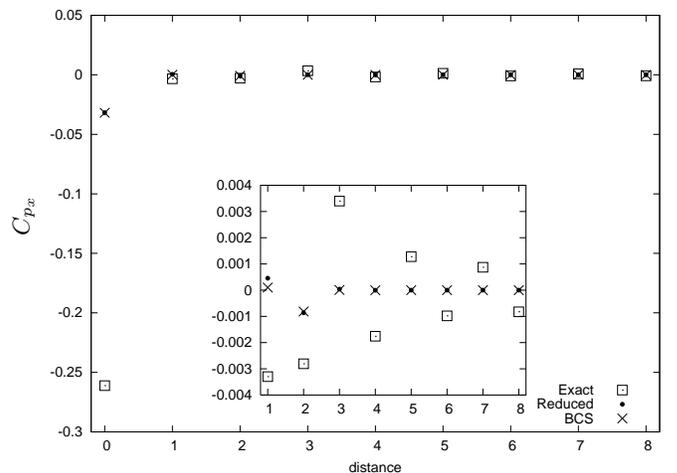}
\caption{The $p$-wave correlators as a function of distance (one dimension,
$N_e=16$).}
    \label{fig:P1_16}
  \vsp
\end{center}
\end{figure}

\begin{figure}[htb]
  \begin{center}
        \vspt
   \epsfxsize=23pc \epsfbox{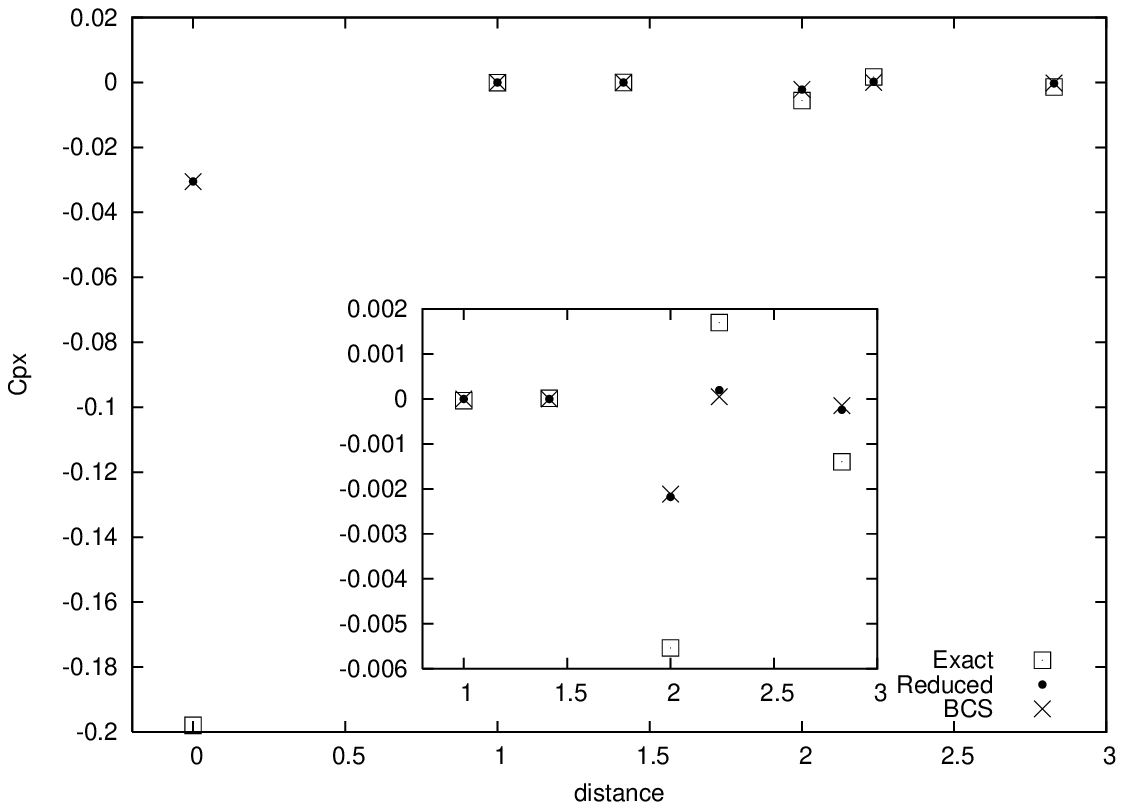}
\caption{The $p$-wave correlators as a function of distance (two dimensions,
$N_e=16$).}
    \label{fig:P2_16}
  \vsp
\end{center}
\end{figure}

\begin{figure}[htb]
  \begin{center}
        \vspt
   \epsfxsize=23pc \epsfbox{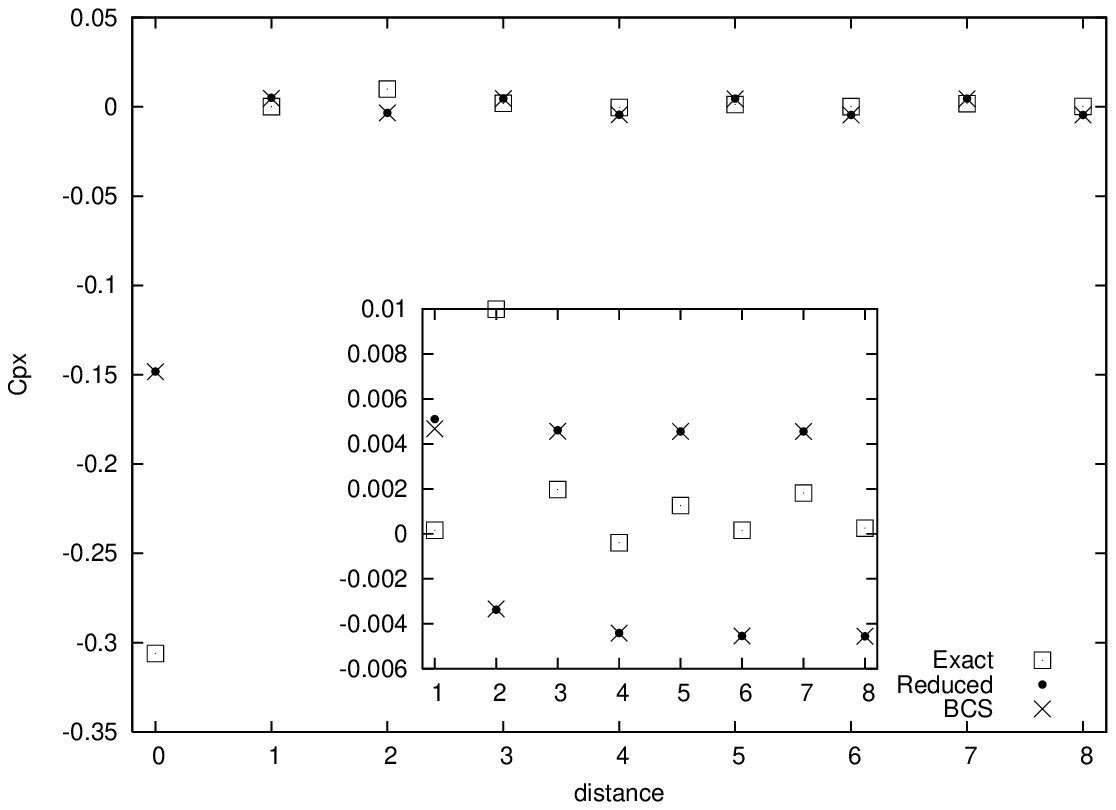}
\caption{The $p$-wave correlators as a function of distance (one dimensions,
$N_e=15$).}
    \label{fig:P1_15}
  \vsp
\end{center}
\end{figure}

\begin{figure}[htb]
  \begin{center}
        \vspt
   \epsfxsize=23pc \epsfbox{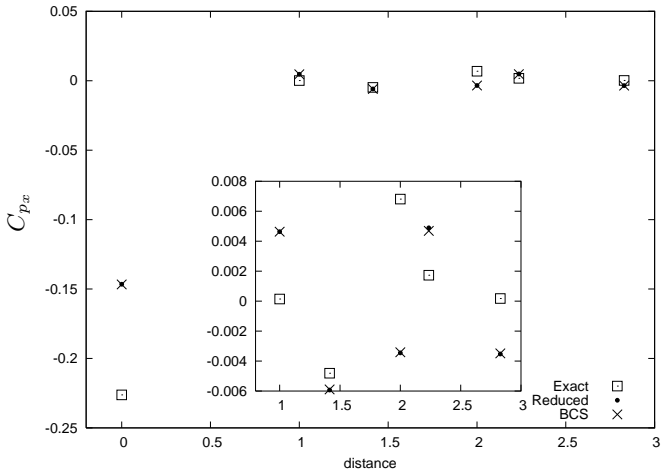}
   \caption{The $p$-wave correlators as a function of distance (two dimensions,
$N_e=15$).}
    \label{fig:P2_15}
  \vsp
\end{center}
\end{figure}

The $p$-wave correlators, $C_{p_x}(d)$, are plotted and compared
in Figs. \ref{fig:P1_16}, \ref{fig:P2_16}, \ref{fig:P1_15} and \ref{fig:P2_15}
for different fillings and dimensions. Both interesting and somewhat
surprising
is that the differences between the BCS and reduced correlators are
now discernable, especially in 2 dimensions.  Since these two model
wavefunctions are identical to within 1 part in $10^3$, this indicates
that the $p$-wave correlator is a very sensitive wavefunction probe as
this correlator's matrix
elements are dominated by a very small portion of the state vector.  This
result also
reflects the small $p$-wave correlator magnitude.  None of the
models exhibit
strong long range order or significant $p_x$ correlations.
Finally, the differences  between Hubbard and reduced
$p$-wave correlators are dramatic for both one and two dimensions with even
or odd filling.

\begin{figure}[htb]
  \begin{center}
        \vspt
   \epsfxsize=23pc \epsfbox{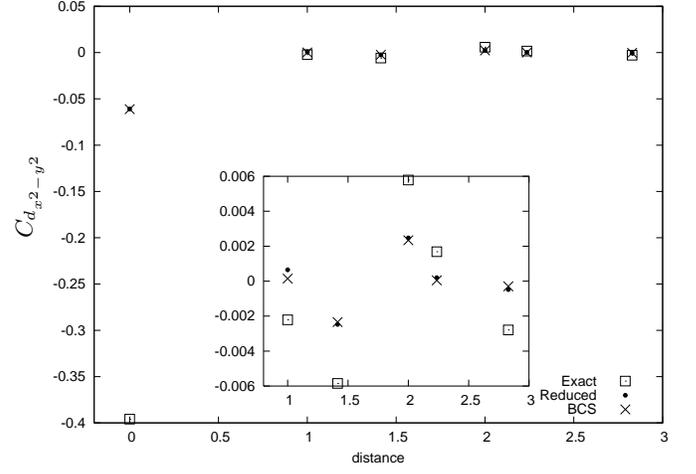}
\caption{The $d$-wave correlators as a function of distance (two dimensions,
$N_e=16$).}
    \label{fig:D2_16}
  \vsp
\end{center}
\end{figure}

\begin{figure}[htb]
  \begin{center}
        \vspt
   \epsfxsize=23pc \epsfbox{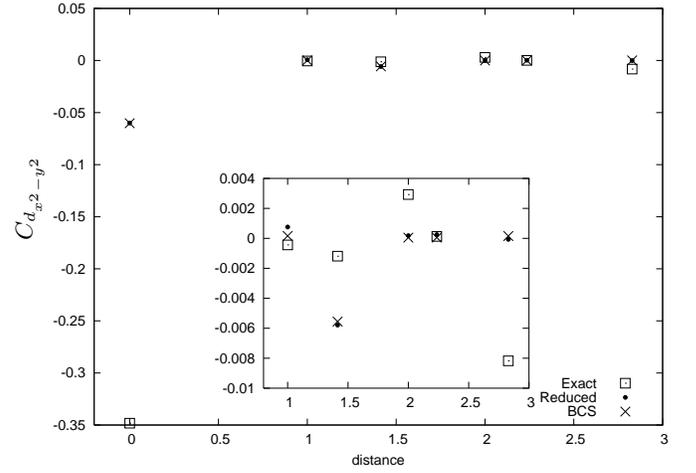}
\caption{The $d$-wave correlators as a function of distance (two dimensions,
$N_e=15$).}
    \label{fig:D2_15}
  \vsp
\end{center}
\end{figure}

Finally, the $d$-wave correlators, $C_{d_{x^2 - y^2}}$, were
calculated in two dimensions (Figs. \ref{fig:D2_16} and
\ref{fig:D2_15}).  As with the $p$-wave correlators, the magnitudes
are small for non-zero $d$ and the differences between the Hubbard and
reduced correlators are dramatic.  The differences between the reduced
and BCS correlators are again present.

\FloatBarrier
\section{Conclusions}

The major thrust of this work has been to further numerically assess
the accuracy of the BCS approximation.  By applying this approximation
to multi-electron systems governed by both the Hubbard and a simpler,
reduced Hamiltonian, each exactly diagonalizable, it has been
documented that the BCS provides an excellent description of ground
state energies in both one and two dimensions.  However, the BCS single
particle energy gaps are less accurate,
in both one and two dimensions, especially near half filling.
This undermines previous assertions that the BCS
approximation improves with increasing dimension.  Further,
propagators, overlaps and correlators have been found
to be quite  sensitive to both interactions and approximations as
the BCS and
reduced model predictions are appreciably different than the Hubbard
results.  The slow decrease of the Hubbard and reduced
Hamiltonian ODLRO propagators and
correlator functions with increasing lattice separation is evidence for
superconductivity.
However correlations, and thus the degree of superconductivity, are much
stronger in the
reduced model. Finally, the BCS method extremely accurately reproduces all
reduced model observables.

In summary, our results challenge the validity of the BCS method as a good,
comprehensive
approximation to the attractive Hubbard model.  While  the Hubbard
and BCS correlators both exhibit long-range order,  their variations with
distance are quite different.  Because agreement between
correlators and inner-product overlaps improves to some extent with
dimension,  perhaps
in three dimensions the
BCS method may yet be a very accurate alternative to exact Hubbard model
wavefunctions but further
study is necessary.   Finally, for physical systems
where the interactions embodied  in the reduced Hamiltonian dominate,
the BCS approximation will be a very accurate approximation scheme permitting
significantly more ambitious numerical investigations.

\begin{acknowledgments}
  The authors would like to thank Dean Lee, Frank Marsiglio, 
  Lubos Mitas, Greg Recine and Thomas
  Schaefer for informative and
  fruitful communications. 
  Computer time provided by Applied Electronics Lab, Stevens Institute
  of Technology is also appreciated.  This work was supported in part
  by grants DOE DE-FG02-97ER41048 and NSF DMS-0209931.

\end{acknowledgments}


\begin{thebibliography}{99}



\bibitem{bcs}
J. Bardeen, L.N. Cooper and J.R. Schrieffer, Phys. Rev. {\bf 108}, 1175 (1957).

\bibitem{bog}
N.N Bogoliubov, Nuovo Cimento {\bf 7}, 794 (1958).

\bibitem{val}
J.G. Valatin, Nuovo Cimento {\bf 7}, 843 (1958).

\bibitem{dmp}
K. Dietrich, H.J. Mang and J.H. Pradal, Phys. Rev. {\bf 135}, {B\bf 22} (1964).

\bibitem{flsc}
F.J. Llanes-Estrada and S.R. Cotanch, Phys.\ Rev.\ Lett.\ {\bf 84}, 1102
(2000).

\bibitem{mar}
F. Marsiglio, Phys. Rev.  {B\bf 55}, 575 (1997).

\bibitem{bd}
F. Braun and J. von Delft, Phys.\ Rev.\ Lett.\ {\bf 81}, 4712 (1998).

\bibitem{tmar}
K. Tanaka and F. Marsiglio, Phys. Rev.  {B\bf 60}, 3508 (1999).

\bibitem{fmh}
W. Fettes, I. Morgenstern and T. Husslein, Int. J. Mod. Phys. {C\bf 8},
1037 (1997).

\bibitem{lw}
E.H. Lieb and F.Y. Wu, Phys. Rev. Lett. {\bf 20}, 1445 (1968).

\bibitem{schrieffer}
J.R. Schrieffer, {\it Theory of Superconductivity} (W.A. Benjamin
Inc., New York, NY 1964).

\bibitem{merminwagner}
N.D. Mermin and H. Wagner, Phys.\ Rev.\ Lett.\ {\bf 17}, 1133 (1966)
and D. Ghosh, Phys.\ Rev.\ Lett.\ {\bf 27}, 1584 (1971)
proved that expectation value is zero for spin wave operators
in the Heisenberg and Hubbard models, respectively.
D. Jasnow and M.E. Fisher, Phys. Rev.  {B\bf 3}, 907 (1971) 
proved, by the same methods, that a nonzero ODLRO
is explicitly forbidden in the Heisenberg model.  

\bibitem{solvable} The high U expansion of the Bethe ansatz is given
  by M. Takahashi, {\it Thermodynamics of one-dimensional solvable
    models} (Cambridge University Press, Cambridge, UK 1999).  
  By symmetry, $N_{e\uparrow} = N_{e\downarrow}$, with $U
  < 0$, is  equivalent to $N_{e\uparrow} + N_{e\downarrow} = V$,
  with $U>0$.

\end{thebibliography}
\end{document}